\documentclass[fleqn,usenatbib]{mnras}

\usepackage[T1]{fontenc}

\usepackage{multirow} 

\DeclareRobustCommand{\VAN}[3]{#2}
\let\VANthebibliography\thebibliography
\def\thebibliography{\DeclareRobustCommand{\VAN}[3]{##3}\VANthebibliography}

\usepackage{graphicx}	
\usepackage{amsmath}	
\usepackage{amssymb}	
\usepackage{threeparttable}
\def\beq{\begin{equation}}
\def\eeq{\end{equation}}
\def\bey{\begin{eqnarray}}
\def\eey{\end{eqnarray}}
\def\Myr{\, {\rm Myr} }
\def\Gyr{\, {\rm Gyr} }

\def\pc{\, {\rm pc} }

\def\kpc{\, {\rm kpc} }

\def\msun{M_\odot}

\def\kms{\, {\rm km \, s}^{-1} }

\def\grad{{\bf \nabla}}

\def\gext{{\bf g}_{\rm ext}}
\def\gint{{\bf g}_{\rm int}}
\def\tdyn{{\rm t}_{\rm dyn}}

\def\Q0{Q_0}

\def\rvir{R_{\rm vir}}

\title[CRGs formation in MD]{Formation of collisional ring galaxies in Milgromian dynamics}
\author[L. Ma and X. Wu]{
Li Ma,$^{1,2}$
Xufen Wu$^{1,2}$\thanks{Corresponding author: xufenwu@ustc.edu.cn}
\\ 
$^{1}$CAS Key Laboratory for Research in Galaxies and Cosmology, Department of Astronomy, \\~~University of Science and Technology of China, Hefei, 230026, P.R. China
\\
$^{2}$School of Astronomy and Space Science, University of Science and Technology of China, Hefei 230026, P.R. China\\
}

\begin{document}
\label{firstpage}
\pagerange{\pageref{firstpage}--\pageref{lastpage}}
\maketitle

\begin{abstract}
  Ring galaxies are rare in the Universe. A head-on or off-centre collision between an intruder galaxy and a disc galaxy leads to a collisional ring galaxy (CRG) when the intruder-to-target mass ratio (ITMR) is over $0.1$ in Newtonian dynamics. Within the framework of Milgromian dynamics, the strong equivalence principle is violated due to the presence of an external field. When two galaxies collide towards each other, the dynamical mass of the phantom dark halo in a galaxy is suppressed by the external field induced by the other galaxy. As a consequence of such suppression, the gravitational perturbation for the target galaxy introduced by a collision is weakened. In this case, a minor collision may not be capable of generating a CRG. In this work, we address this issue and perform a series of numerical simulations of collisions by tuning the values of ITMR. We find that the critical ITMR is $0.5$ in MOND, which is much larger than that in Newtonian dynamics. The observed massive ring galaxies, such as Arp 147, can be effectively interpreted by CRGs in MOND. This interpretation does not necessitate the presence of dark matter halos for either the target or intruder galaxies. Moreover, for a larger inclination angle or a larger impact parameter, the off-centred ring structure is fainter. The larger critical ITMR indicates that it is harder to form a CRG in Milgrom's Modified Newtonian Dynamics (MOND). To account for the observed ring structures of the NGC 922-like galaxies in MOND, it is necessary to invoke other scenarios than a dry minor collision.
\end{abstract}

\begin{keywords}
galaxies: interactions - galaxies: kinematics and dynamics - gravitation - methods: numerical 
\end{keywords}



\section{Introduction}
A ring galaxy was identified by \citet{Zwicky1941}, followed which more ring galaxies have been observed \citep{Vorontsov-Velyamino1959,Arp1966,deVaucouleurs1975,Theys_Spiegel1977,Madore+2009,Conn+2011}. The ring structures in galaxies could be generated by several possible mechanisms, including the Lindblad resonances induced by galactic bars \citep{Buta_Combes1996,Buta+1999}, acceration and merger of galaxies \citep{Schweizer+1983,Bekki1998}, and collisions between a disc target galaxy and an intruder galaxy \citep{Lynds_Toomre1976,Theys_Spiegel1976}. Most ring galaxies are products of galaxy collisions \citep{Few_Madore1986,Hernquist_Weil1993,Appleton_Struck-Marcell1996}, thus the collisional ring galaxies (CRGs) are natural laboratories to examine the interactions of galaxies. 

Within the framework of Newtonian dynamics, there are numerous existing numerical simulations to investigate the formation of CRGs with different parameters of initial conditions, including the orbital parameters such as initial relative velocity \citep{Fiacconi+2012}, impact parameter \citep{Toomre1978,Fiacconi+2012}, inclination angle \citep{Lynds_Toomre1976,Ghosh_Mapelli2008}, and mass ratio between the target and the intruder galaxies \citep{Hernquist_Weil1993,Horellou_Combes2001}, and also including the gas fraction \citep{Mapelli_Mayer2012}, the bulge-to-disc mass ratio \citep{Chen+2018} and the Toomre instability of the target galaxy \citep{Guo+2022}. In these simulations, a ring forms in a disc galaxy after an off-centre or head-on collision with a dwarf galaxy, if the dynamical mass of the dwarf galaxy is larger than $10\%$ of the target galaxy. In Newtonian simulations, both galaxies are embedded in cold dark matter (CDM) halos.

Milgrom's MOdified Newtonian dynamics \citep[MOND, or Milgromian dynamics,][]{Milgrom1983a}, an alternative to dark matter halo models, proposed that a modification to gravitation may explain the apparent dark matter effect \citep{Milgrom1986a}, namely phantom dark matter. When the gravitational acceleration is below a critical value of $3.7 ~{\rm km}^2{\rm s}^{-2} \pc^{-1}$, denoted as $a_0$, the gravitational acceleration of a system decreases as $r^{-1}$ by taking into account the above modification. The mass discrepancy on scales from star clusters of a few parsecs to local supervoid and the Hubble tension of a few Gpc can be well explained by MOND \citep[e.g., see the reviews of][]{Sanders_McGaugh2002,Famaey_McGaugh2012,Banik_Zhao2022}. Apart from the natural prediction of the baryonic Tully-Fisher relation \citep{TF1977}, MOND accounts for various contexts from the small scale of a few thousand au to the large scale of clusters of galaxies. These include the recently observed gravitational anomaly in the wide binaries \citep{Chae2023a,Hernandez2023} at the solar neighbourhood. While \citet{Banik+2024} posited that the kinematic accelerations of wide binaries within $250pc$ of the sun have ruled out MOND at $16\sigma$ confidence, \citet{Chae2023b} presented a contrasting perspective. According to \citet{Chae2023b}, there is a noticeable deviation from Newtonian dynamics for wide binaries with a separation exceeding 2 kau. This deviation, significant at about $5\sigma$, lends support to MOND. On the scale of globular clusters (GCs) of the Milky Way, several observations and simulations have suggested deviations from Newtonian dynamics. This is evidenced by the line-of-sight (LoS) velocity dispersion profiles of these GCs remain flat at large radii \citep{Scarpa+2007,Scarpa+2011,Lane+2009,Lane+2010,Scarpa_Falomo2010,Hernandez_Jimenez2012,Durazo+2017}. The overall LoS velocity dispersion for an isolated and isotropic self-consistent system in MOND is $\sigma_{\rm LoS}^2=\frac{2}{9}(GM_ba_0)^{1/2}$ with $M_b$ being the baryonic mass \citep{Milgrom1994}. Thus the flat LoS velocity dispersion profiles can be naturally reproduced in MOND.  MOND appears to offer a compelling explanation of the kinematics and dynamics on the scales of GCs, as well as the central \citep{Milgrom2009} and outer regimes \citep{Kroupa+2012,Kroupa2012,Kroupa2015} of galaxies. Despite reproducing the flat rotation curves and the baryonic Tully-Fisher relation, a radial acceleration relation (RAR) has been observed in galaxies \citep{McGaugh2004,McGaugh+2016,Lelli+2016}, which is predicted by MOND \citep{Milgrom1983a}. Recently, the RAR has been extended to weak-lensing data from the fourth data release of the Kilo-Degree Survey \citep[KiDS-1000,][]{Brouwer+2021}, and an agreement between MOND prediction and the weak-lensing data has been found.

On large scales, MOND encounters several challenges. For instance, there may be a need for additional dark matter at the scale of clusters of galaxies. This is suggested by lensing data from clusters \citep{Angus+2007,Natarajan_Zhao2008}, the Bullet Cluster 1E0647-56 \citep{Clowe+2006} and the dark matter ring surrounding the galaxy cluster Cl0024+17 \citep{Jee+2012}. Moreover, MOND seems to predict an incorrect amplitude for the mass function of galaxy clusters \citep{Angus_Diaferio2011,Angus+2013,Angus+2014}. The third peak of the angular power spectrum of the cosmic microwave background (CMB) was once a challenge for MOND \citep{Skordis+2006}. However, with the introduction of a new relativistic MOND theory by \citet{Skordis_Zlosnik2021,Skordis_Zlosnik2022}, the CMB temperature and polarisation angular power spectra no longer pose a problem for MOND. This advancement has significantly enhanced the explanatory power of MOND in the context of CMB spectra. Moreover, the matter power spectrum (MPS) obtained from the Sloan Digital Sky Survey (SDSS) data release 7 (DR7) \citep{Reid+2010} can be perfectly reproduced in this relativistic MOND.

In the context of MOND, a uniform and constant external field, $\gext$, changes the enhancement of internal gravitation for a self-bound system \citep{BM1984,Milgrom1986a,Famaey+2007,Wu+2008,Haghi+2009}. That is, the strong equivalence principle (SEP) breaks down \citep[e.g.,][]{Milgrom1983a,BM1984,Chae+2020}.  One of the detectable external field effects is the lopsided shapes of stellar systems embedded in constant background gravitational fields \citep{Wu+2010,Wu+2017,Candlish+2018}. The shapes of these systems are asymmetric along the near and far sides of the directions of the external field sources. Such asymmetric structures can also be observed in the tails of globular cluster streams \citep{Thomas+2017} and open clusters \citep{Kroupa+2022}. The strength of the external field at the solar neighbourhood of our Galaxy is approximately $1.6-2.0~a_0$ \citep{Blanchet_Novak2011,Kroupa+2022}. This suggests that local open clusters are in the mild-MOND regimes and are predominantly influenced by the external field, as the internal gravitational acceleration (denoted as $\gint$) is much smaller than $\gext$. In nearby open clusters orbiting the Milky Way, the numbers of stars in the leading and trailing tails exhibit asymmetry in MOND due to the external field \citep{Kroupa+2022}. 

Furthermore, the internal acceleration of a system declines in the presence of an external field, and stars with high mechanical energy are enabled to escape from the system \citep{Milgrom1983a,Famaey+2007,Wu+2007}. The phantom dark matter halo is thus truncated at a radius, denoted as $\rvir$, for a system dominated by an external field \citep{Wu_Kroupa2015}. If a system is eccentrically orbiting a massive galaxy, the strength of the external field is spatially varying. Hence $\rvir$ and the truncation mass of the phantom dark matter halo change at different locations along the orbit. For instance, in an outgoing orbit, the external field from the massive galaxy evolves from strong to weak, which results in a phase transformation for the system from quasi-Newtonian to Milgromian \citep{Wu_Kroupa2013}. The phantom dark matter mass of the system is increasing on such an orbit \citep{Wu_Kroupa2015}. Most previous works focused on internal systems transfer from quasi-Newtonian to Milgromian dynamics \citep[e.g.,][]{Wu_Kroupa2013,Haghi+2016,Haghi+2019,Famaey+2018,Kroupa+2018,Haghi+2019}. However, the collisions of galaxies, in the opposite direction of phase transformation, are yet to be studied. In an incoming orbit of a collision, the mass of the phantom dark matter halo for an intruder galaxy becomes smaller. Therefore the gravitational perturbation introduced by the intruder galaxy is expected to be weaker than that in Newtonian dynamics. Although CRGs can be naturally produced by galaxy collisions in Newtonian dynamics, it is unclear whether a collisional ring could form in MOND. In principle, the intensity of the perturbation is enhanced by the increasing mass of the intruder galaxy. Thus a key parameter for the formation of CRGs is the ITMR, i.e., the intruder-to-target mass ratio in a collision of two galaxies. 

In this work, we will perform a systematic study on the formation of CRGs within the framework of MOND. We will explore the new critical mass of ITMR to generate a CRG. The numerical models and the colliding orbits are described in \S \ref{MONDmodels}. To obtain a more generic conclusion, we perform the simulations of collisions on both head-on (\S \ref{ITMR}) and off-centre (\S \ref{offcentre}) orbits in MOND for massive, intermediate and low-mass galaxies. We summarise our results in \S \ref{conclusions}.

\section{MOND and numerical models}\label{MONDmodels}
The modification to gravitational acceleration in the original version of MOND \citep{Milgrom1983a} follows
\beq \label{original}
{\bf g}_N=\mu(X){\bf g},~~~~~X=\frac{g}{a_0},
\eeq
where ${\bf g}_N$ and ${\bf g}$ are the Newtonian and Milgromian acceleration generated by the same gravitational source, and $g=|{\bf g}|$. The function $\mu(X)$ is an interpolating function that satifies when $X \gg 1.0$, $\mu \rightarrow 1.0$, and when $X \ll 1.0$, $\mu \rightarrow X$. The original version is simple to use, but is only accurate when a self-bound system is a point mass, a spherically symmetric, or a cylindrically symmetric system.

To conserve the energy, momentum and angular momentum, a self-consistent Lagrangian version of MOND, so-called AQUAL, was proposed by \citet{BM1984}. In a self-bound system, the amount of phantom dark matter is a $100\%$ conspiracy with baryons, following the modified Poisson's equation of \citep{BM1984,Famaey+2007}
\bey\label{poisson}
-\nabla &\cdot& \left[ \mu(X) ({\bf g}_{\rm ext} - {\mathbf\grad} \Phi_{\rm int}) \right]=4\pi G \rho_{\rm b},\\
\qquad X&=&{|{\bf g}_{\rm ext} - {\mathbf\grad} \Phi_{\rm int}| \over a_0} .\nonumber
\eey
In the above equation, $\Phi_{\rm int}$ is the internal potential introduced by the baryonic density $\rho_{\rm b}$, and ${\bf g}_{\rm ext}$ is the external gravitational acceleration. There are several popular forms for the interpolation function $\mu(X)$, among which the ``standard'' form, $\mu(X)=X/(1+X^2)^{1/2}$, fits best for the circular velocities of spiral galaxies \citep{Sanders_McGaugh2002}, and the ``simple'' form, $\mu(X)=X/(1+X)$, fits best for the Milky Way \citep{Famaey_Binney2005}. In practice, solving the non-linear Poisson's equation, Eq. \ref{poisson}, is hard in AQUAL. The numerical simulations take a much longer CPU time to solve Eq. \ref{poisson}, compared to Newtonian Poisson's equation.

To benefit from the above two forms of MOND (i.e., Eqs \ref{original} and \ref{poisson}), a new, QUasi-linear formulation of MOND (QUMOND) was developed by \citet{Milgrom2010}, in which the authors derived the formulation from action, and thus the conservation laws are satisfied. The modified Poisson's equation writes
\beq\label{QUMOND}
\nabla^2 \Phi =\nabla \cdot [\nu(Y)\nabla \Phi_{\rm N}], ~~~~Y=|\nabla \Phi_{\rm N}|/a_0,
\eeq
where $\Phi$ and $\Phi_{\rm N}$ are the Milgromian and Newtonian potentials, respectively, and $\nu (Y)=1/ \mu (X)$. The $\nu$ function approaches 1 when $Y\gg 1$ and tends to become $1/\sqrt{Y}$ when $Y \ll 1$. The two popular forms of the interpolating $\nu$ function, corresponding to the $\mu$ function in AQUAL, are
\bey\label{nu}
\nu(Y) &= &\left(0.5+0.5\left(1+4/Y^2\right)^{1/2}\right)^{1/2},~~~~{\rm ``standard"},\nonumber \\
\nu(Y) &=& 0.5+\sqrt{0.25+1/Y}, ~~~~{\rm ``simple"},
\eey
respectively. We shall adopt the ``simple'' form in this work. Eq. \ref{QUMOND} can be rewritten in the form of
\beq\label{linear}
\nabla^2 \Phi = 4\pi G(\rho_{\rm b} +\rho_{\rm PDM}),
\eeq
where $\rho_{\rm PDM}$ is the phantom dark matter mass density, and $\rho_{\rm PDM}= \frac{1}{4\pi G} \nabla \cdot [(\nu -1) \nabla \Phi_{\rm N})]$. As the right-hand side of Eq. \ref{linear} is only relevant to the Newtonian potential $\Phi_{\rm N}$, the formulation is easier to solve compared to Eq. \ref{poisson}. Thus a rapid N-body simulation within the framework of MOND is possible. QUMOND differs from AQUAL with a curl field, but the two formulations are distinguishable in spherically symmetric systems \citep{Milgrom2010,Zhao_Famaey2010}. \citet{Candlish2016} compared the two formulations of AQUAL and QUMOND in structure formation simulations by using a parallel adaptive mesh refinement code, {\it RAYMOND} \citep{Candlish+2015}, and found that the results closely resemble each other.

We shall perform the simulations by using an adaptive mesh refinement code, {\it Phantom of RAMSES} \citep[hereafter {\it PoR,}][]{Lueghausen+2015,Nagesh+2021}, which modifies Poisson's equation in the code {\it RAMSES} \citep{Teyssier2002}. In {\it PoR}, the QUMOND formulation, i.e., Eq. \ref{QUMOND}, is computed at each timestep.
  
\subsection{Galaxy models}
In the present work, we aim to study the formation of CRGs within the framework of MOND. The dominant mechanism to causes the expected distinct differences in the CRGs between the standard Newtonian dynamics and MOND is gravitation. Thus we will perform pure N-body simulations with different initial conditions in MOND, and compare our results with the currently existing studies on the CRGs in the framework of Newtonian dynamics. Both the target and intruder galaxies are live particle systems in this work.

\subsubsection{Target galaxy models}
The observed CRGs are mostly massive galaxies similar to the Milky Way \citep[e.g.,][]{Gerber+1992,Marcum+1992,Conn+2011,Elagali+2018}. Until now, the lowest-mass CRG system with an overall mass of stars and HI gas of $6.6\times 10^9\msun$ is observed in the Galactic neighbourhood \citep{Parker+2015}, Kathryn's Wheel (ESO 179-13), which is a triple system with Components A, B and C. Component C is neglected in the interaction of the colliding system since the mass is too small. The stellar masses for the other two galaxies, Components A and B, are $1.3\times 10^9\msun$ and $1.7\times 10^8\msun$. The HI gas for Component A is unknown, but the overall HI for the galaxy-pair system is $2\times 10^9 \msun$.

In our models, the target galaxy is a disc following an exponential and a sech$^2$ distribution along the radial and vertical directions \citep{Hernquist1990},
\beq \rho_{\rm d}(R,z)=\frac{M_{\rm d}}{4 \pi h^2 z_0} \exp \left(-\frac{R}{h}\right) {\rm sech}^2\left(\frac{z}{z_0}\right).\eeq
To make a systematic investigation of the modified gravity effect, the values for the overall stellar mass of the target galaxy, $M_{\rm{d}}$, are desired to be $1.072\times 10^{11}\msun$, $1.072\times 10^{10}\msun$ and  $1.072\times 10^{9}\msun$, corresponding to the mild-, moderate- and deep-MOND cases, respectively. The masses we choose for the target model represent the masses from the Milky-Way-like disc galaxies to the Kathryn's-Wheel-like galaxies.
$h$ and $z_0$ are the radial and vertical scale lengths for the disc galaxy models, and their values are shown in Table \ref{models}.

\begin{table*}\centering
\caption[]{The parameters of the mass density distributions for the target galaxies ($1_{st}-4_{th}$ columns) and the intruder galaxy models corresponding to each target galaxy ($7_{th}$ and $8_{th}$ columns).  $f$ is the mass ratio between an intruder and a target galaxy. The values of $f$ are chosen as 0.1, 0.3, 0.4, 0.5, 0.6, 0.8, and 1.0. The dynamical time scales for the target galaxy models are listed in the $5_{th}$ column. The internal accelerations at the position where the rings are expected to be observed are displayed in the $6_{th}$ column.
}
\begin{tabular}{cccccc|cc}
  \hline 
  &\multicolumn{5}{c|}{Target galaxy} & \multicolumn{2}{c}{Intruder galaxy}\\
  Number of particles&\multicolumn{5}{c|}{ $4\times 10^6$} & \multicolumn{2}{c}{$1\times10^5$}\\
  \hline
  Model & $M_{\rm d}$ & $h$  & $z_0$  & $\tdyn$  & $g(R=3h)$& $M_{\rm i}$ &$r_{\rm p}$\\
  & $(10^{10}\msun)$ & $(\kpc )$ &$(\kpc)$ &$(\Myr)$&$(a_0)$  &$(10^{10}\msun)$ & $(\kpc)$ \\
   \hline
  T1 & 10.72 & 4.0 &0.8   &$77.5$& 1.3 &$f\times 10.72$&0.75 \\
  T2 & 1.072 & 1.5 &0.3   &$51.7$&1.0  &$f\times 1.072$ &0.35 \\
  T3 & 0.172 & 1.0 &0.2  &$61.3$& 0.4 &$f\times 0.1072$&0.15\\
  \hline
\end{tabular}\label{models}
\end{table*}

To generate the initial conditions (ICs) for the disc galaxy models, a modified version \citep{Banik+2020a,Nagesh+2021} of the {\it Disk Initial Conditions Environment (DICE)} code \citep{Perret+2014,dice-code} is used. The adapted version of {\it DICE} calculates the superpositions of particles by a Metropolis-Hasting Monte-Carlo Markov Chain algorithm using the QUMOND gravitational acceleration ${\bf g}$. The acceleration follows ${\bf g} =\nu {\bf g}_{\rm N}$ in a spherical approximation, where a ``simple'' $\nu$ function is adopted. There is an inaccuracy caused by the curl field $\nabla \times {\bf g}$ in a disc system. To avoid the inaccuracy, a modification to the $g_{\rm Nz}$ in Eq. \ref{nu} is taken into account, which is $g_{\rm Nz}=2\pi G\Sigma \tanh(2)$ for a thin disc \citep{Banik+2020a}. Here $\Sigma$ is the surface density of the disc, and $G$ is the gravitational constant. The Toomre parameter, $Q$, is generalised to QUMOND in the form of \citep{Banik+2018}
\beq
Q\equiv \frac{\sigma_{\rm r}\Omega_{\rm r}}{3.36 G' \Sigma},
\eeq
with $G'=G\nu \left(1+\frac{1}{2}\frac{\partial \ln \nu}{\partial \ln g_{\rm N}}\right)$, where $\sigma_{\rm r}$ and $\Omega_{\rm r}$ are the radial velocity dispersion and radial epicyclic frequency, respectively. The value of $Q_{\rm lim}=1.5$ is set in the modified version of {\it DICE}, to ensure $Q\ge 1.5$ everywhere. Hence the ICs for the disc are supposed to be stable.  The number of particles for a disc galaxy model is $4\times 10^6$. 

A cold thin disc is dynamically unstable in Newtonian dynamics\citep{Binney_Tremaine1987}. A bar forms in a self-gravitating pure disc due to the instability \citep{Miller+1970,Hohl1971}. The disc can be stabilised by a central compact stellar bulge \citep{Sotnikova_Rodionov2005}, or by a dark matter halo with a concentrated centre \citep{Toomre1981,Sellwood_Evans2001}. In MOND, however, the situation is different. \citet{Milgrom1989} found that a thin disc is more stable when the self-gravitation is weak, i.e., the disc is dominated by a deep MOND gravity. Numerical simulations later confirmed that a disc is more stable for a decreasing average acceleration, ${\bf g}$, generated by the disc \citep{Brada_Milgrom1999}, since the distribution of the phantom dark matter in a disc galaxy precisely follows that of baryons. The stability of a disc galaxy in MOND is out of the scope of this work, and we will not discuss this too much. 

To avoid additional evolution caused by the gravitational instability during the formation of ring galaxies, we freely evolve all the disc models for $1~\Gyr$ using {\it PoR} before performing the simulations of collisions. The free evolution tests and the following simulations of collisions are performed in a box with a size of $2500~\kpc$. The simulation units in {\it PoR} are $l_{\rm simu}=1.0~\kpc$, $M_{\rm simu}=10^{10}\msun$ and $G=1.0$. Thus the simulation time unit is $T_{\rm simu}=4.7~\Myr$. The minimal and maximal levels of refinements are $l_{\rm min}=8$ and $l_{\rm max}=19$, respectively. The actual maximal level of refinement in the simulations is $l_{\rm max}^{\rm acut}=17-19$ for models in Table \ref{models}, corresponding to a maximal spatial resolution of $19.5- 4.8~\pc$. We provide an example of a free evolution test in the appendix, \S \ref{stability}. A dynamical time is defined as
\beq \tdyn=\frac{r_{90\%}}{v_{c,90\%}},\eeq
where ${v_{c,90\%}}\approx (GM_{90\%}a_0)^{1/4}$ is the MOND circular velocity at a radius enclosing $90\%$ mass of a target galaxy. This radius is denoted as $r_{90\%}$, and $M_{90\%}$ is the inner $90\%$ mass of a model. For models $T1$, $T2$ and $T3$, the values of $\tdyn$ are approximately $77.5~\Myr$, $51.7~\Myr$ and $61.3~\Myr$, respectively. The time scale for the free evolution is at least $13$ dynamical times for all the target galaxy models, thus long enough to reach their stable state (see the appendix \S \ref{stability}). 
The target galaxies models are the final products of the freely evolved discs.

\begin{figure}
  \includegraphics[width=90mm]{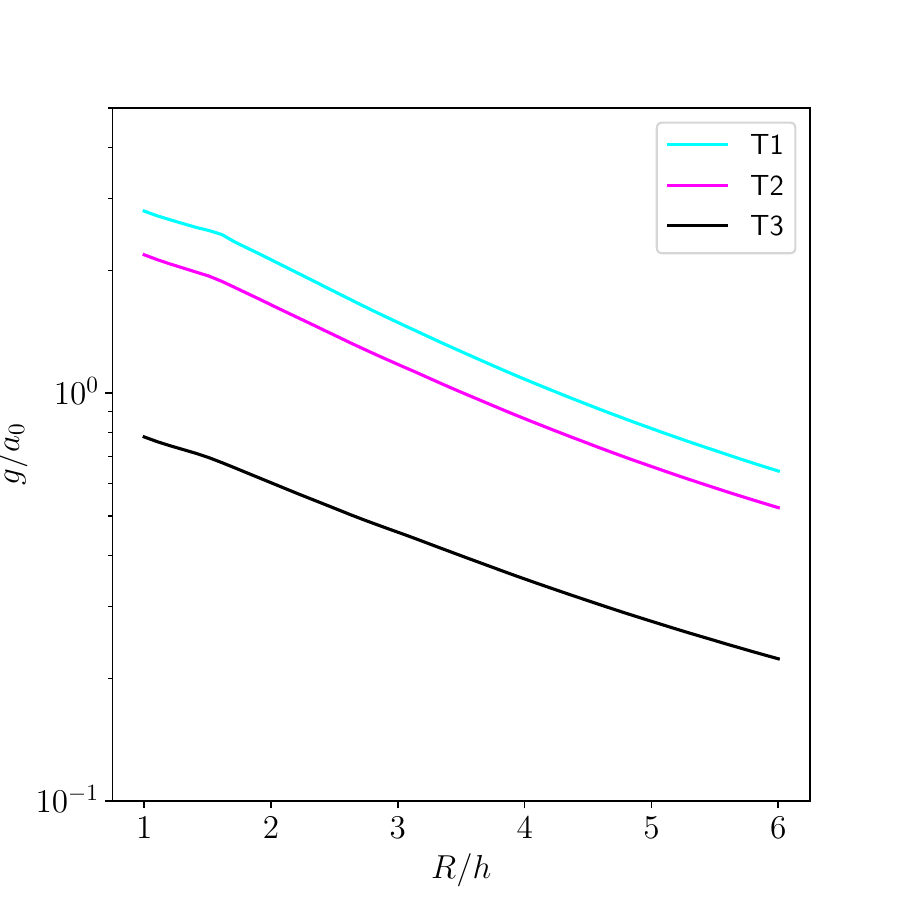}
  \caption{The internal acceleration of the target models $T1$, $T2$ and $T3$ in units of $a_0$.
  }
  \label{gint}
\end{figure}

In addition, we present the internal gravitational acceleration of the target galaxy models in Fig. \ref{gint}. For the three models, $T1$, $T2$ and $T3$, the internal accelerations at $R=h$ are approximately $2.8 a_0$, $2.2 a_0$ and $0.8a_0$, respectively. In the outer regions where the rings are anticipated to be observed (for instance, at $R=3h$), the corresponding internal accelerations are $1.5a_0$, $1.2a_0$ and $0.5a_0$. Consequently, the internal accelerations in the outer regions of these three models are within mild- ($g> a_0$), moderate-  ($g\approx a_0$) and deep-MOND ($g < a_0$) gravity regimes.

\begin{table*}\centering
\caption[]{The table provides the parameters for the collisional models. The $f1$-$f10$ models are head-on collisions for a mild-MOND target galaxy. The $moderate1$, $moderate2$ and $deep1$, $deep2$ models are for the moderate- and deep-MOND cases with different ITMR. The $moderate6$-$moderate10$ rows list the moderate-MOND models with tuning values of $b$ at a fixed ITMR of 1.0. 
The bottom rows are the off-centre collision sets with different values of ITMR, $b1$-$set$ and $b2$-$set$. The $2_{\rm nd}$-$4_{\rm th}$ columns list the ITMR, the initial positions and relative velocities of the intruder galaxy in galactocentric coordinates of the target galaxy. The masses of the target and intruder galaxies are shown in the $5_{\rm th}$ and $6_{\rm th}$ columns, respectively. The $7_{\rm th}$ column lists the values of impact parameter $b$, and the corresponding inclination angle $\theta$ is provided in the $8_{\rm th}$ column. The masses of rings propagating to radii of $3h$ and $5h$ are listed in the $9_{th}-10_{th}$ columns. Finally, the fractions of mass loss for the intruder galaxy in the collisions are listed in the $11_{th}$ column.}
\begin{threeparttable}
\setlength{\tabcolsep}{1mm}{
\begin{tabular}{ccccccccccccc}
  \hline
  Model & $f$&$(x_{\rm 0},~y_{\rm 0},~z_{\rm 0})$ & $(v_{\rm x0},~v_{\rm y0},~v_{\rm z0})$  & $M_{\rm d}$ & $M_{\rm i}$ & $b$ & $\theta$ & $M_{\rm ring}^{3h}$&$M_{\rm ring}^{5h}$ & $M_{\rm loss}$\\
 & & $(\kpc)$ & $(\kms)$ & $(10^{10}\msun)$ & $(10^{10}\msun)$ & $(\kpc)$ & $(^\circ)$ & $(10^{10}\msun)$ & $(10^{10}\msun)$ & (\%)\\
 \hline
  $f1$ &0.1&$(0.0,~0.0,~-48.2)$ & (0,~0,~600) &10.72 & 1.07 & 0.0 &0.0 & - & -&4.8\\
  $f3$&0.3&$(0.0,~0.0,~-48.2)$ & (0,~0,~600)  &10.72 & 3.22 & 0.0&0.0 & - & - &3.2\\
  $f4$&0.4&$(0.0,~0.0,~-48.2)$ & (0,~0,~600) &10.72 & 4.29 & 0.0&0.0 & - & - & 2.3\\
  $f5$&0.5&$(0.0,~0.0,~-48.2)$ & (0,~0,~600) &10.72 & 5.36 & 0.0&0.0 & 2.83& -& 2.3\\
  $f6$&0.6&$(0.0,~0.0,~-48.2)$ & (0,~0,~600) &10.72 & 6.43 & 0.0&0.0 &4.03 & -& 1.8\\
  $f8$&0.8&$(0.0,~0.0,~-48.2)$ & (0,~0,~600) &10.72 & 8.58 &0.0&0.0 &6.16&3.50 &1.1\\
  $f10$&1.0&$(0.0,~0.0,~-48.2)$ & (0,~0,~600)&10.72 &10.72&0.0&0.0 & 6.21&5.27&0.7\\
$moderate1$&0.6 &$(0.0,~0.0,~-18.1)$ & (0,~0,~190) &1.07 & $6.43\times 10^{-1}$& 0.0&0.0& 0.23 &-&2.2\\
$deep1$ &0.6&$(0.0,~0.0,~-12.1)$ & (0,~0,~60)  &$1.07\times 10^{-1}$ & $6.43\times 10^{-2}$& 0.0&0.0 & - & - &0.0\\
$moderate2$&1.0 &$(0.0,~0.0,~-18.1)$ & (0,~0,~190) &1.07 & 1.07& 0.0&0.0 & 0.53 & 0.37&1.3\\
$deep2$ &1.0&$(0.0,~0.0,~-12.1)$ & (0,~0,~60)  &$1.07\times 10^{-1}$ & $1.07\times 10^{-1}$& 0.0&0.0 & 0.04 & - & 0.0\\
$moderate6$&1.0 &$(-10.1,~0.0,~-18.1)$ & (95,~0,~190) &1.07 & 1.07& 0.9&26.6 & 0.56 & 0.37 & 1.1\\
$moderate7$&1.0 &$(-12.2,~0.0, ~-18.8)$ & (111,~0,~190) &1.07 & 1.07& 1.1& 30.3& 0.55 & 0.34 &1.0\\
$moderate8$&1.0 &$(-14.4,~0.0,~-19.5)$ & (126,~0,~190) &1.07 & 1.07& 1.2& 33.7 & 0.51 & 0.31 &0.9\\
$moderate9$&1.0 &$(-16.9,~0.0,~-20.3)$ & (142,~0,~190) &1.07 & 1.07& 1.4&36.9 & $0.48\tnote{c}$ & $0.28\tnote{c}$ &0.7 \\
$moderate10$&1.0 &$(-19.5,~0.00,~-21.1)$ & (158,~0,~190) &1.07 & 1.07& 1.5&39.8 & $0.45\tnote{c}$ &$0.25\tnote{c}$ &0.7 \\
\hline
\multicolumn{10}{c|}{Off-centre collisions with varying ITMR: $0.1\le f \le 1.0 $} \\
\hline
$b1$-$set$ && $(-4.00,~0.00,~-43.35)$ & (50,~0,~600) &10.72 & $f \times 10.72$&0.40 &4.75 & & & &\\
$b2$-$set$ && $(-8.10,~0.00,~-43.80)$ & (100,~0,~600)&10.72& $f \times 10.72$ & 0.80& 9.46 & & & &\\
   \hline
\end{tabular}}\label{orbpara}
\begin{tablenotes}\footnotesize
\item[c] Here a C-shaped structure forms instead of a completed ring. The value is the mass of the C-shaped structure.
\end{tablenotes}
\end{threeparttable}
\end{table*}

\subsubsection{Intruder galaxy models}
For each target galaxy, a set of collision simulations is performed with different mass models for the intruder galaxy. The mass models for one set of these intruder galaxies follow Plummer's density profile \citep{Plummer1911}, which is given by 
\beq\label{Toomre}
\rho(r)=\frac{3M_{\rm i}}{4\pi r_{\rm p}^3}\left(1+\frac{r^2}{r_{\rm p}^2} \right)^{-5/2},
\eeq
where $M_{\rm i}$ is the stellar mass of the intruder galaxy, and $M_{\rm i}=f\times M_{d}$. Here $f$ is the ITMR. The values of $f$ are chosen as $0.1$, $0.3$, $0.4$, $0.5$, $0.6$, $0.8$ and $1.0$, covering from the minor collisions to the major collisions. The watershed for minor and major collisions is $f=0.3$, as commonly used in the existing works \citep[e.g.,][]{Elagali+2018b}.
 $r_p$ is the Plummer scale length. The parameters for each set of intruder galaxy models are listed in Table \ref{models}. For each intruder galaxy model, there are $10^5$ particles. The particle ICs for the intruder galaxy are generated by {\it McLuster} code \citep{Kuepper+2011} within the Newtonian framework, and they have been converted into a Milgromian system by multiplying $\sqrt{-W/2K}$ to the velocities of all particles, where $W$ and $K$ are the Clausius integral in a MOND system and the kinetic energy of the Newtonian ICs, respectively. The system is thus re-virialised in MOND gravitation. Once the live particle models for the colliding galaxies are successfully prepared, the colliding galaxy pairs are placed on a series of starting points of orbits in Cartesian coordinates with the centre of the target galaxy being the origin. We describe the parameters for the orbits in the following subsection and Table \ref{orbpara}.

\subsection{Orbits for the colliding galaxy pairs}
\begin{figure*}
  \includegraphics[width=190mm]{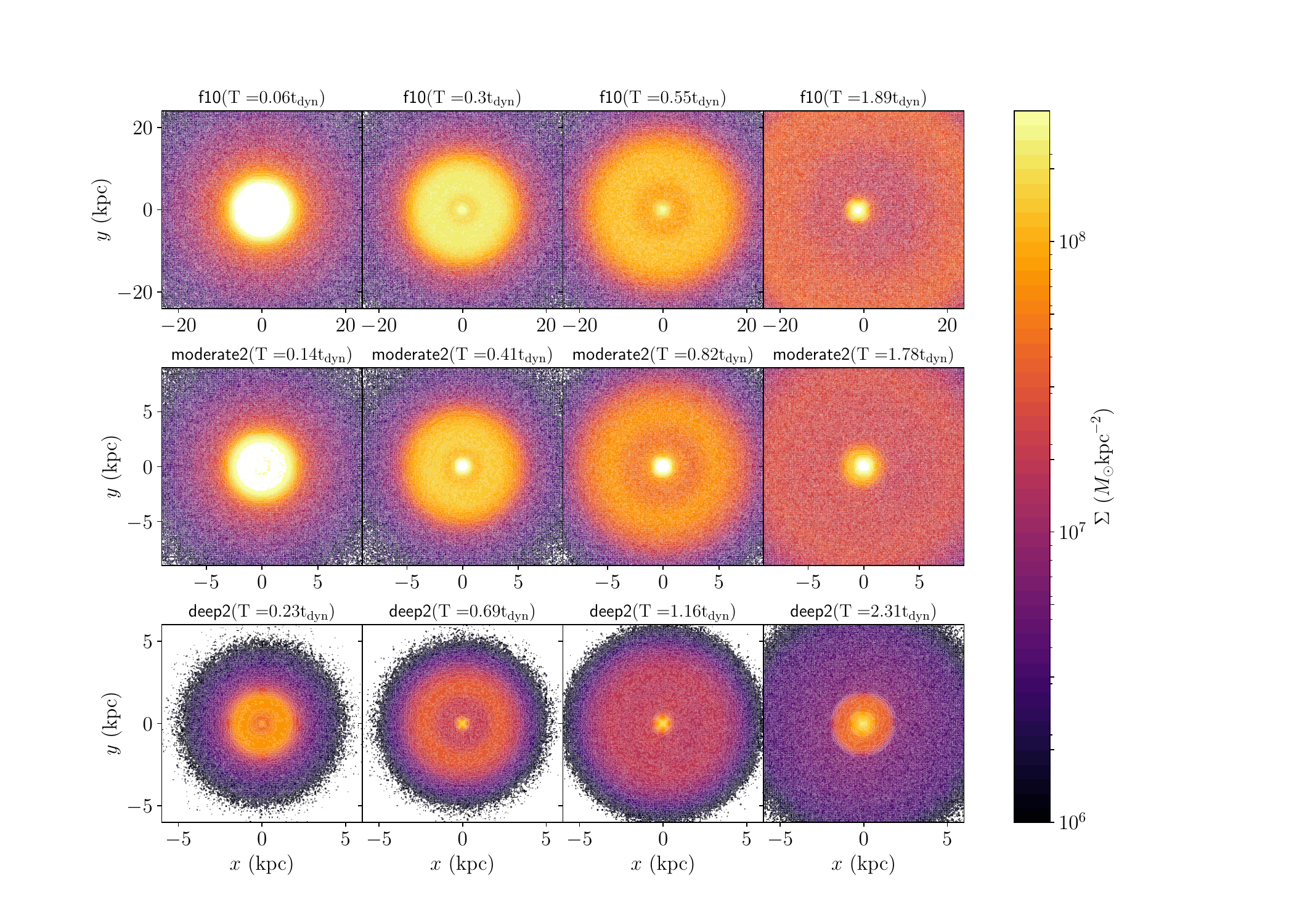}
  \caption{The projected density maps illustrate the evolution of collisional ring structures with different target galaxy models. The panels, arranged from top to bottom, display the target galaxies $T1$, $T2$ and $T3$ undergoing head-on collisions with an ITMR of 1.0. These correspond to mild-, moderate- and deep-MOND cases. The different evolution time scales are labelled above each panel in units of $\tdyn$. The leftmost panels show the projected densities of discs at the moment when the intruder galaxies go through the disc planes. The second column of panels demonstrates the emergence of rings. The third column of panels displays the propagation of these rings out to a radius of $R=3h$. The rings disappear at large radii and are illustrated in the rightmost panels. 
  }  \label{ringevo}
\end{figure*}

In one collision, the initial positions of the intruder galaxy are $(x_{\rm 0},~y_{\rm 0},~z_{\rm 0})$, and the initial relative velocities are $(v_{\rm x0},~v_{\rm y0},~v_{\rm z0})$ in the galactocentric coordinates of the target galaxy. The intruder galaxy collides with the target galaxy along the vertical direction of the latter one. The schematic diagrams of the orbits are the same as that in \citet{Chen+2018}. Our simulations include both head-on and off-centre collisions. The parameters of the collisional orbits for the intruder galaxy in our simulations are listed in Table \ref{orbpara}. Firstly, we study the formation of CRGs on a head-on orbit by tuning the ITMR, and the corresponding simulations are $f1$-$f10$ models in Table \ref{orbpara}. After successfully obtaining a CRG from a collision, we change the mass of the target galaxy at two given mass ratios to represent the gravity from mild- to deep-MOND. They are models $f6$, $moderate1$ and $deep1$ for a fixed value of $f=0.6$ and models $f10$, $moderate2$ and $deep2$ for another fixed value of $f=1.0$. 
Furthermore, we choose the mild-MOND models to change the impact parameter, $b$, of the intruder galaxy to non-zero values, which stand for the off-centre collisions (the $b1-set$ and the $b2-set$ models). Finally, we perform a set of off-centre collisions in the moderate-MOND gravity regimes. These models maintain a fixed ITMR of $f=1.0$, while adjusting the value of $b$ to determine the maximum value that still allows for the formation of a ring (the $moderate6$-$moderate10$ models).

\begin{figure*}
  \includegraphics[width=190mm]{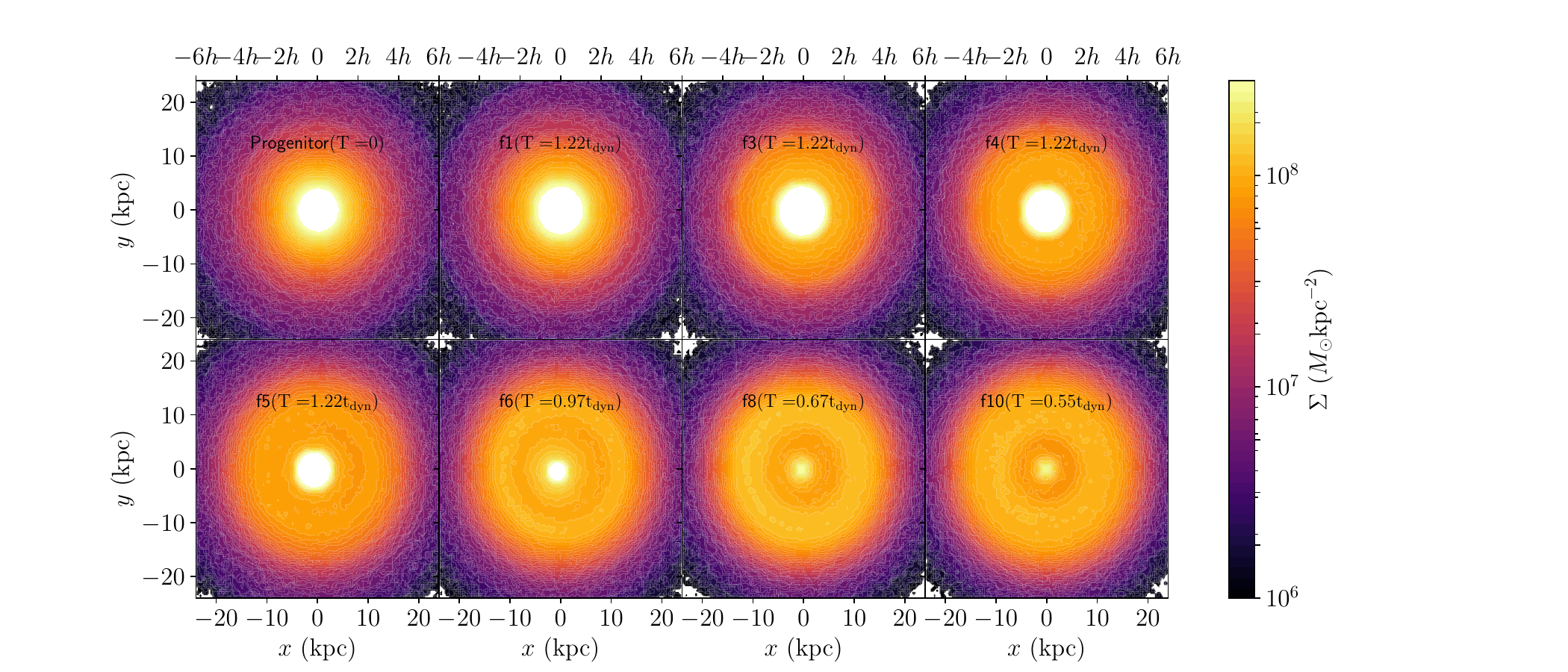}
  \caption{The projected density maps of the disc planes undergone head-on collisions with different values of ITMR $f$. No rings form when $f<0.5$.
  }  \label{ITMR2d}
\end{figure*}

\section{Head-on collisions}\label{ITMR}
As aforementioned, an external field truncates a phantom dark matter halo at $\rvir$. The truncation radius is where the strengths of the external and internal accelerations are comparable \citep{Wu_Kroupa2013,Wu_Kroupa2015}. Thus a stronger external field leads to a smaller $\rvir$. The gravitational field from a target galaxy dampens the mass of the phantom dark matter halo of an intruder galaxy. For a dwarf galaxy that approaches a disc plane along the vertical direction, the dwarf galaxy is embedded in a spatial-varying external gravitational field induced by the disc galaxy. The external field becomes stronger on such an orbit, and thus the mass of the phantom dark matter halo of the dwarf galaxy evolves to smaller \citep{Wu_Kroupa2013}. In such a process, the gravitational perturbation on the disc plane induced by the dwarf galaxy becomes weaker. However, in Newtonian dynamics, the situation is quite different. In Newtonian dynamics, the change of halo mass of an intruder galaxy is negligible in a flyby collision. Hence the integral gravitational perturbation applied on the disc plane tends to be much more robust in Newtonian on the same orbit. Thus a question arises as to whether a head-on collision, in which the perturbation caused by the dwarf galaxy is much weaker, gives rise to the formation of a collisional ring in MOND.

\subsection{Mild-MOND CRGs}
Since the ITMR is a crucial parameter in forming a ring structure, the first set of simulations mainly focuses on tuning ITMR. We follow the definition of ``ring formation'' by \citet{Chen+2018}, a ring structure emerges at $x=+2h$ and propagates further than $+2h$ means the formation of an observable ring through a collision. The existing time scale, $t_{\rm exist}$, is defined as the duration between the formation at $+2h$ and the disappearance at the maximal propagation radius,  denoted as $r_{\rm max}$, of a ring. We demonstrate three examples of the evolution of CRGs in different MOND gravity regimes in Fig. \ref{ringevo}. The examples presented involve different mass models for the target galaxies, namely $T1$, $T2$ and $T3$ (as shown from the top to bottom panels of Fig. \ref{ringevo}). The corresponding intruder galaxies in these examples all have an ITMR of $1.0$. The orbital parameters for the three collisions are described as the models $f10$, $moderate2$ and $deep2s$ in Table \ref{orbpara}. This figure presents the projected density maps of the target galaxies at various stages of interaction with the intruder galaxy. From left to right, we display the moments when the intruder galaxies go through the disc planes, the emergence of rings, the propagation of rings to a radius of $R=3h$ and the eventual disappearance of the rings. The time scales are labelled above each panel in units of $\tdyn$. It is observed that a ring tends to be thinner and fainter when it propagates to $R=3h$ as the target galaxy mass decreases. This intriguing phenomenon warrants further investigation. In the following sections, we will delve deeper into the distinct evolution of rings within different MOND gravity regimes.

Fig. \ref{ITMR2d} shows the projected density distributions of the disc planes when rings propagate to $R=3h$ in the $f1$-$f10$ models. No rings propagate out to $3h$ with an ITMR of $f<0.5$ after two galaxies go through each other. We plotted the projected density profiles of those failed CRG models at the same time scale of a successful CRG model, model $f6$, when its ring structure propagates out to $R=3h$. In the successful CRG model, a ring propagates to $R=3h$ on the disc plane $75\Myr \approx 1.0\tdyn$ after the two galaxies go through each other.

\begin{figure}
  \includegraphics[width=90mm]{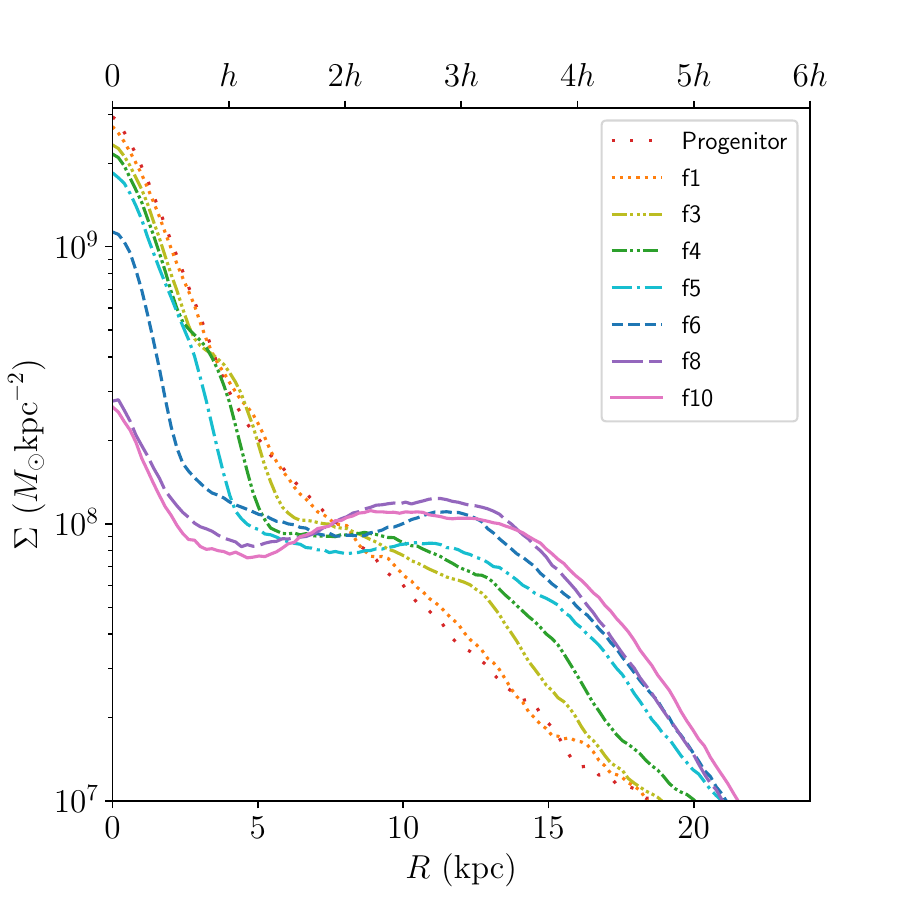}
  \caption{The projected mass density of the progenitor (red curve) and of the discs (models $f5$-$f10$) for different values of ITMR at the time scale of the rings propagating to $R=3h$. For models that do not form rings (models $f1$-$f4$), the curves display the projected density at the same time scale of the $f6$ model with a ring at $3h$.
  }
  \label{ITMR1d}
\end{figure}

\subsubsection{Failed CRGs at $3h$ with low ITMR values}
We carefully examine the failed CRG models at $3h$ with low ITMR values. For the model $f1$, only an inner ring structure forms after the collision and disappears in the inner region of the disc within $1h$. Such an ``inner ring'' is not considered to be a CRG, as the propagation distance is too small and thus hard to observe. For the model $f3$, an inner ring structure emerges at $R=2h$ but vanishes at a radius smaller than $3h$. When the first ring vanishes, the stellar particles are still moving radially outwards. There appears to be an expanding extended platform on the disc plane. A faint ring structure at $R=3h$ is generated when the propagation of the extended platform turns radially inwards due to the disc gravity approximately $250\Myr \approx 3\tdyn$ after the collision. Since the ring structure is too faint and is not the first ring, we do not consider it a typical CRG. When the ITMR increases to 0.4, the situation is very similar to that in model $f3$. The above results are very different to those in the standard Newtonian models \citep[e.g.,][]{Hernquist_Weil1993,Horellou_Combes2001,Chen+2018}. In Newtonian gravity, the critical ITMR to form a ring is 0.1 and the ring-like structure can propagate further than $5h$. The reason for the failed CRGs at $3h$ with low ITMR values is that the enhancement of gravity in MOND follows exactly the distribution of baryonic matter on the disc \citep[see Fig. 3 in][]{Wu+2008}. A disc with a given Toomre's $Q$ parameter is more stable in MOND than that in Newtonian gravity \citep{Milgrom1989,Brada_Milgrom1999,Banik+2020a}. Thus the stabilised disc is harder to form a ring-like structure and also harder to propagate to a large distance if the perturbation from a collision is not strong enough.

\subsubsection{Successful CRGs at $3h$ and further distances}
\begin{figure}
  \includegraphics[width=90mm]{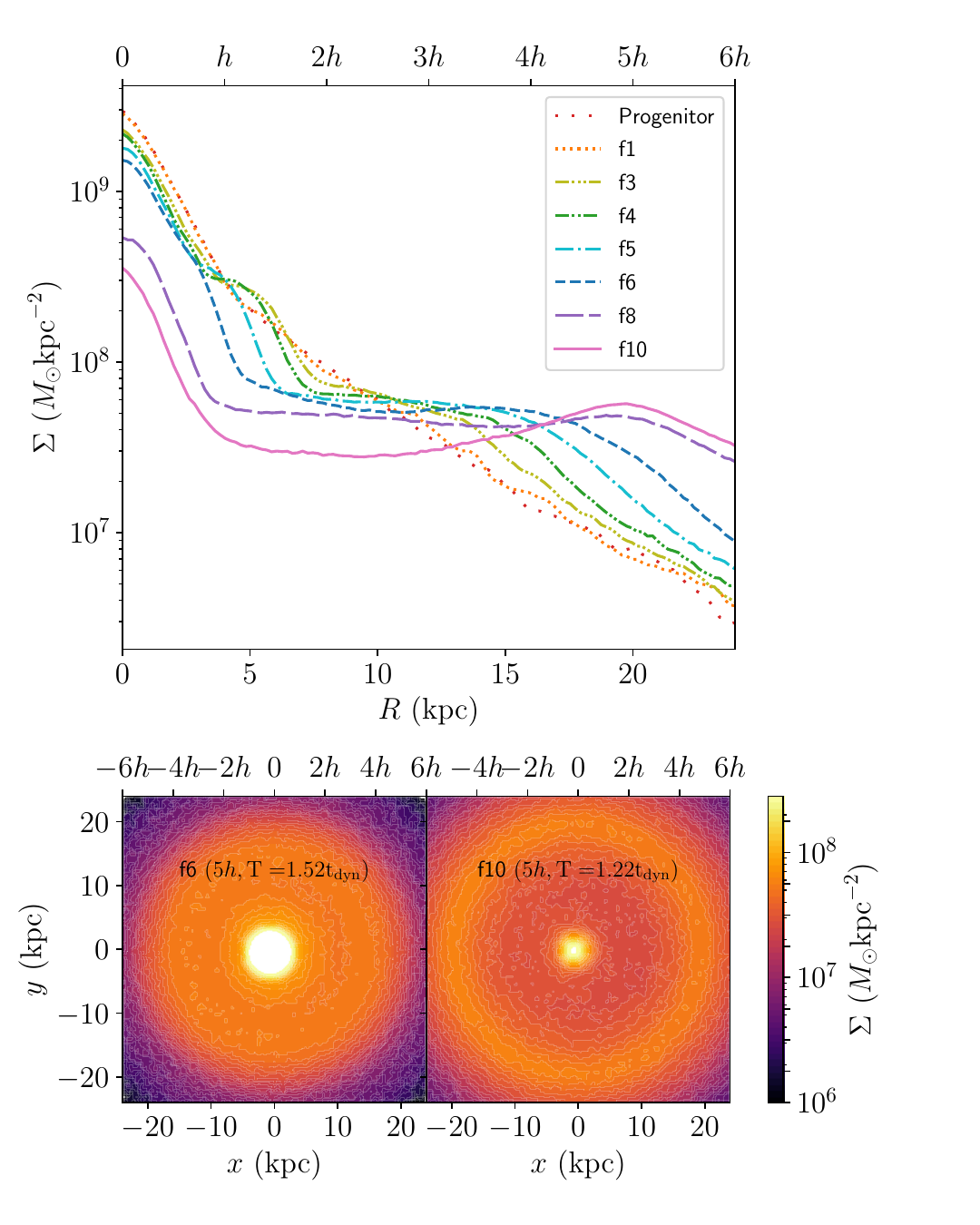}
  \caption{The upper panel shows the projected density profiles $\Sigma(R)$ of the target galaxy at the time scale of a peak propagating to $5h$ for models $f8$ and $f10$. The $\Sigma(R)$ profiles of galaxies that do not form a ring at $5h$ are also demonstrated at the same time scale of model $f8$ using different colours indicated in the panel. The lower two panels display the projected density maps for models $f6$ (lower-left) and $f10$ (lower-right) at these time scales.
  }
  \label{vanish}
\end{figure}

When $f=0.5$, a faint and cloudy ring structure forms on the disc plane and propagates radially outwards. When the ITMR increases, the ring structures become clearer. The 1-dimensional column density profiles for ring galaxies, when the rings propagate to $R=3h$, are plotted in Fig. \ref{ITMR1d}, together with the failed-CRG-models at the same time scale of the $f6$ model, i.e., $75 \Myr\approx 1\tdyn$ after the collisions. For a model, the contrast between the peak at $R=3h$ and the valley inside $3h$ presents the intensity of the ring structures. There are no peaks near $3h$ for models $f1$, $f3$ and $f4$. A peak appears at $3h$ for the model $f5$, but the density contrast between the valley and the peak is lower than $10\%$ of the density at the bottom of the valley. For the $f6$ model, the ring structure tends to be clear in Fig. \ref{ITMR2d}. The peak is narrow and clear in Fig. \ref{ITMR1d}. And the density contrast between the peak and the valley is above $10\%$. When the ITMR again increases, the ring becomes thicker, and the corresponding peak at $3h$ is broader, with a larger density contrast. Moreover, the projected density in the central region reduces as the ITMR grows, which implies that more stellar particles move outwards together with the propagation of the ring since the gravitational perturbation caused by a more massive intruder galaxy is more intensive. We note that the ring for model $f10$ at $3h$ is so broad that it is hard to distinguish it from the projected density map (the lower rightmost panel of Fig. \ref{ITMR2d}). The reason is that the perturbation induced by the intruder galaxy is so strong that the ring is still forming at this time scale. When the ring propagates to a larger radius of the galactic centre, the ring becomes thinner and more robust (lower right panel of Fig. \ref{vanish}). The mass of the ring at $3h$ increases as the ITMR grows, since the perturbation from the intruder galaxy tends to be stronger on the same orbit.

The rings continue to propagate outwards after they reach the radius of $R=3h$. For models with smaller values of ITMR, the rings vanish soon. For instance, for model $f6$, the ring structure has disappeared at the radius of $4h$ (Fig. \ref{vanish}). The peak in the projected density profile, $\Sigma(R)$, becomes a platform when the ring disappears. The platform proceeds to move outwards. For model $f10$, a thin and apparent ring structure propagates to $R=5h$ at a later time scale, i.e., $\approx 95 \Myr \approx 1.2\tdyn$ after the collision. We show the projected density profiles $\Sigma(R)$ for the mild-MOND models at the time scale when the rings still exist and propagate out to $5h$ (upper panel of Fig. \ref{vanish}). We find that only in models $f8$ and $f10$ are there density peaks at $5h$. The density peak is already relatively mild for model $f8$. For models that do not form a ring at $5h$, we plot the $\Sigma(R)$ profiles at the same time scale of model $f8$ when its ring structure appears at $5h$. There are no ring signals on the $\Sigma(R)$ profiles for models $f5$ and $f6$, confirming that their rings have vanished. The projected density maps for the two example galaxy models are provided in the lower panels: lower left panel for the $f6$ model and lower right panel for the $f10$ model, respectively. No ring structure can reach up to $5h$ for the $f6$ model. For the $f10$ model, a thin and clear ring can propagate to $5h$. The mass of the ring is approximately half of the stellar mass of the target galaxy. Thus only in a major collision can a ring propagate out to a large radius in the mild-MOND models.

The observed massive ring galaxies, for instance, Arp 147 with a stellar mass of $2.1\times 10^{10}\msun$ \citep {Romano+2008}  for the disc galaxy and a massive intruder galaxy with an ITMR of $1.75$, later the disc galaxy mass is suggested $6.1-10.1\times 10^{10}\msun$ by \citet{Fogarty+2011} and the corresponding ITMR is between $0.6$ and $0.4$, reside in the mild-MOND regions. In the existing Newtonian modellings \citep{Mapelli_Mayer2012}, $f=0.5$ is used by assuming the dynamical mass follows the stellar mass. Such a ring galaxy can be well explained by our collisional scenario in MOND, without the presence of dark matter halos for both galaxies.

\subsubsection{The secondary rings}
\begin{figure}
  \includegraphics[width=85mm]{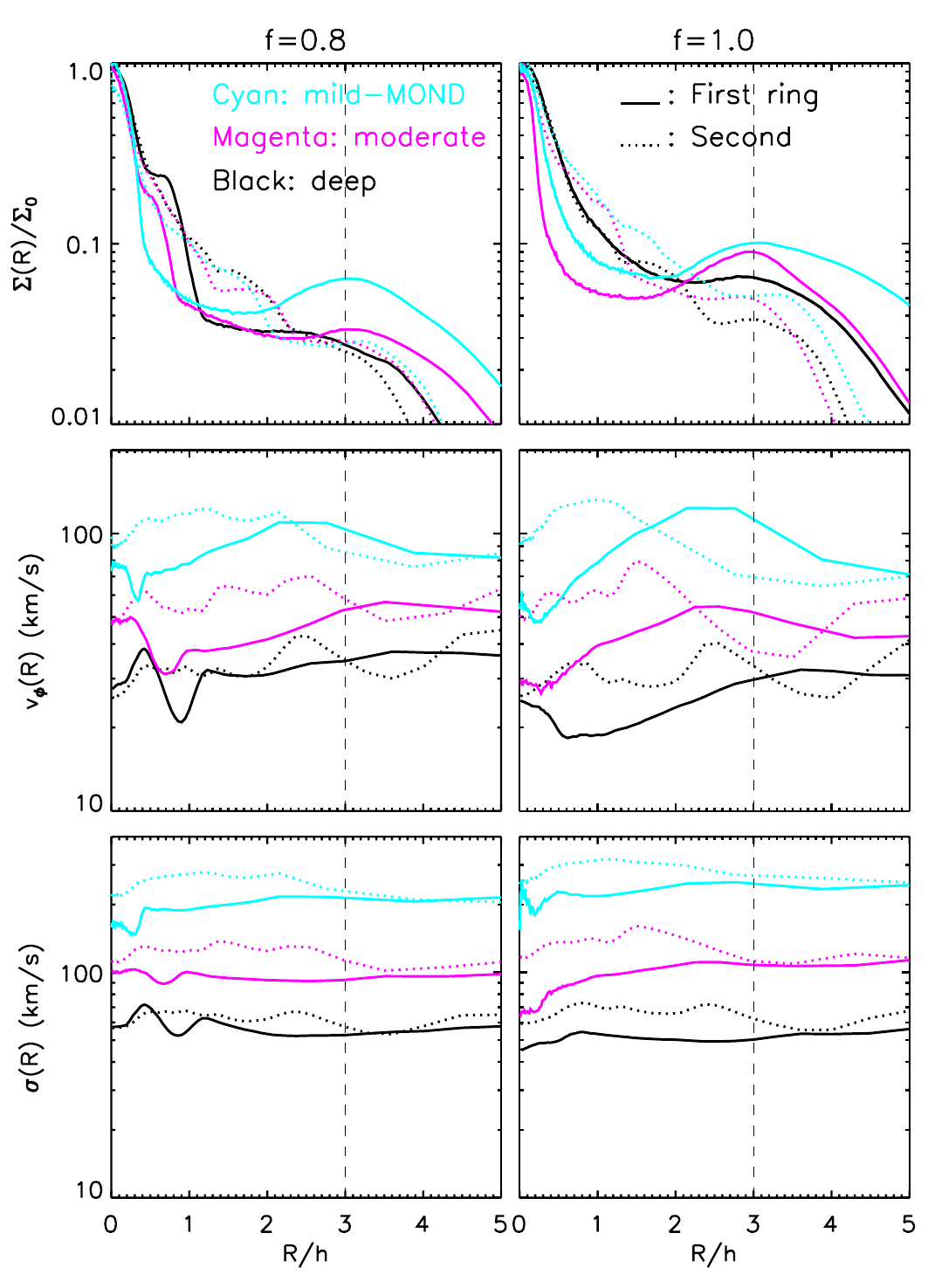}
  \caption{The projected density profiles (upper panels), azimuthal velocity profiles (middle panels) and velocity dispersion profiles (lower panels) for CRG models when the first (solid curves) and second(dotted curves) rings propagate to $3h$. The left panels show the above profiles for models with an ITMR of 0.8 and the right panels display the models with $f=1.0$. Different colours represent different masses for the target galaxy models, cyan for $10^{11}\msun$, magenta for $10^{10}\msun$ and black for $10^{9}\msun$.   }
  \label{secondary}
\end{figure}
There are secondary rings formed in the later stage of the successful CRG models. For the critical ITMR model, model $f5$, the secondary ring is too faint to observe. Instead, an expanding platform on the disc plane can be found after the first ring in-falling back to the disc centre. When the value of ITMR grows, the secondary rings tend to be clearer. A secondary ring emerges at $2h$ for model $f6$ at a time scale of about $175\Myr\approx 2.3\tdyn$ after the collision, but evolves to an out-propagating platform with a constant surface density soon. A faint ring structure at near $5h$ is generated when the platform reaches the maximum radius and falls back. Such a faint ring structure exists for about $30\Myr$ and thus is not expected to be observed in the real Universe. We define the second ring as a ring-like structure formed after the first ring and propagated at least to $3h$.

For a collision with a larger value of ITMR, the formation of the second ring is in the later stage after the collision. In models $f8$ and $f10$, the second rings can naturally form and propagate out to large radii ($3h$ and $5h$). The scenario of a clear primary ring followed by a less developed secondary ring in a head-on collision with a large value of ITMR in mild-MOND gravity is similar to that \citep[e.g.,][]{Gerber+1996,Fiacconi+2012} in Newtonian dynamics. 

After the collision, the second ring of model $f8$ forms at $\approx 260\Myr~(3.4 \tdyn)$, propagates out to $3h$ and vanishes at $5h$ as an extended platform at times scales of about $340\Myr~(4.4\tdyn)$ and $480\Myr~(6.2\tdyn)$. For the model $f10$, a second ring emerges at $380\Myr~(4.9\tdyn)$ and appears at radii of $3h$ and $5h$ at time scales of $\approx 430\Myr ~(5.5\tdyn)$ and $530\Myr ~(6.8\tdyn)$ after the collision. Thus the existing time for the second ring is much longer than that of the first ring. Besides, an increasing ITMR leads to a longer time scale for the formation of the second ring. This can be explained by 
the stronger gravitational perturbation for collisions with a more massive intruder galaxy, such that the more mass of the disc is moving outwards after the collision, and that the existing time scale of the first ring is longer. The first ring reaches its maximal radius and then collapses back to the disc centre. The second ring forms on the disc plane when the stellar material infalls back. 

For the second ring, the contrast of the surface density (the dotted cyan curve) between the peak at $3h$ and the valley ahead of $3h$ is lower compared to the first ring (the solid cyan curve), as shown in the upper panels of Fig. \ref{secondary}. For the first rings of the mild-MOND model, the azimuthal velocity profiles peak at $3h$ (middle panels), in coincidence with the density peaks, while the crests of the second rings fall behind the propagation of the waves of projected densities. The azimuthal velocities are between $50\kms$ and $110\kms$ when the first and second rings propagate to $3h$, which is significantly lower than the circular velocity for a rotation-supported un-perturbed system with a mass of $10^{11}\msun$. The disc is heated by the major collisions, with larger values of the velocity dispersion profiles, as shown in the lower panels of Fig. \ref{secondary}. 

In addition, a very faint third platform-like structure appears to form at $3h$ about $500\Myr~(6.5\tdyn)$ after the collision, and more secondary structures form in the later stages. Since these structures are too faint to observe, we will not focus on them in the CRG systems.

\subsection{Moderate- and deep-MOND CRGs}
\begin{figure}
  \includegraphics[width=90mm]{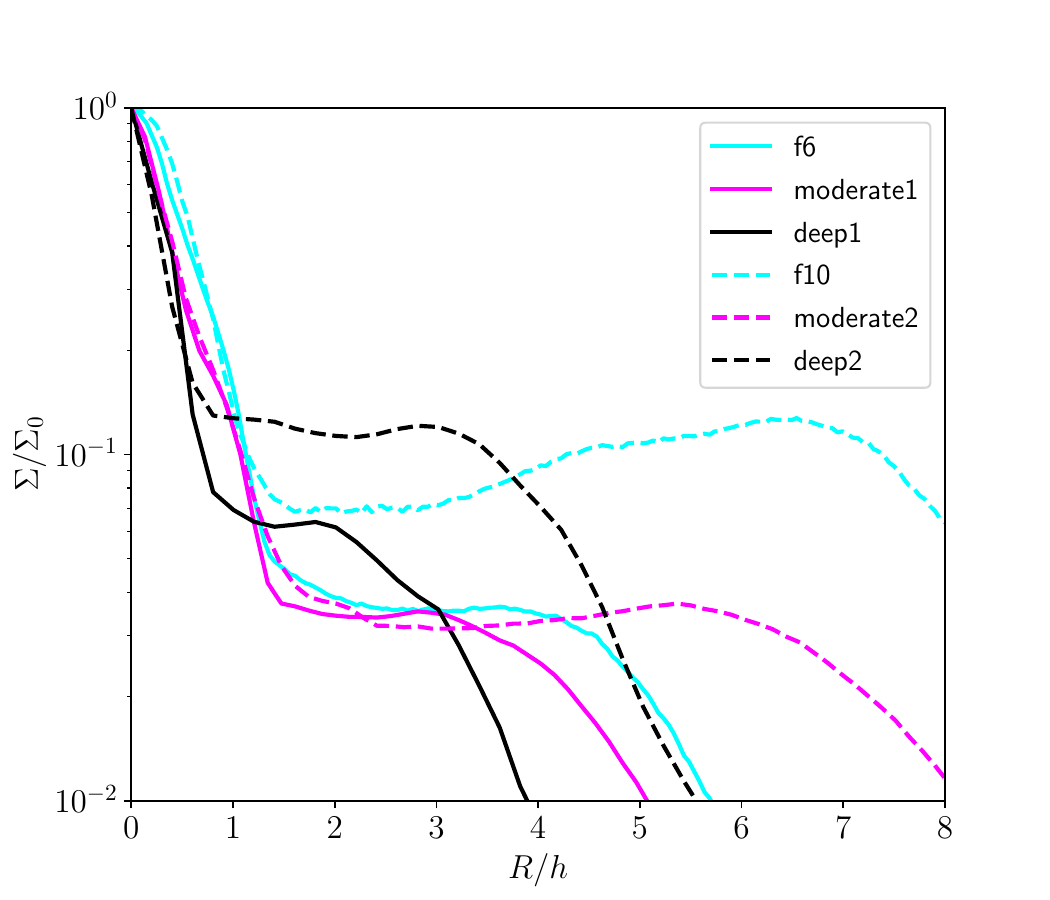}
  \caption{Two sets of the projected density profiles when the rings vanish for models in different MOND-regimes by tuning the mass of the target galaxy. The orange, green and blue curves represent mass models of $1.072\times 10^{11}$, $1.072\times 10^{10}$ and $1.072\times 10^{9}\msun$ for the target galaxy, corresponding to mild, moderate and deep MOND gravities. Different line types display different values of ITMR, solid for $f=0.6$ and dashed for $f=1.0$.
  }
  \label{rvanish}
\end{figure}

For dwarf galaxies dominated by a deep-MOND gravity, the suppression effect of the phantom dark matter caused by an external field is even more severe, such that the existence of CRG is expected to be more challenging in low-mass disc galaxies. To confirm this, we perform simulations with different masses of the target galaxy when the ITMR is fixed at 0.6 and 1.0. The maximum propagation distances of the rings on target galaxies with varying masses are illustrated in Fig. \ref{rvanish}. In a dwarf disc galaxy with a mass of $1.072\times 10^{9}\msun$ with an ITMR of $0.6$ (model $deep1$ in Table \ref{orbpara}), which is in the deep-MOND regime, the ring generated in a collision disappears at a distance of about $2h$. According to the definition of ring formation, $f=0.6$ is the critical ITMR for the low-mass model to form a ring. 

\begin{table*}\centering
\caption[]{The time scales for the first rings of the example CRG models in Table \ref{orbpara}. The dynamical time scales in units of $\Myr$ for each model are shown in the second column. The $3_{rd}$ to $6_{th}$ columns show the time scales of the formation of the rings, the rings expand outwards to $3h$, $5h$ and their maximal propagation radii. The last column shows the existing time scales for the rings.
}
\begin{tabular}{ccccccccccccc}
  \hline
  Model & $\tdyn$ &  $t_{\rm form}$ &  $t(3h)$  &  $t(5h)$ & $t(R_{\rm max})$& $t_{\rm exist}$\\
 &  $(\Myr)$ & $(\tdyn)$ & $(\tdyn)$ & $(\tdyn)$ & $(\tdyn)$ & $(\tdyn)$\\
 \hline
  $f5$&77.5 &0.6 & 1.2 &- & - &1.1 \\
  $f6$&77.5 & 0.4& 1.0&- &1.6 & 1.5\\
  $f8$&77.5 & 0.3& 0.7&1.5 & 1.9 & 1.7\\
  $f10$&77.5 &0.3 & 0.6&1.2 & 1.9 &1.8\\
$moderate1$&51.7  &0.4 & 1.2 &- &1.2 & 1.0\\
$deep1$ & 61.3  & 1.2&- &- &1.2&  0.2\\
$moderate2$&51.7  & 0.4&0.8 & 1.5 & 1.8 & 1.6\\
$deep2$ & 61.3 & 0.7 & 1.2 &- & 1.2& 0.9\\
   \hline
\end{tabular}\label{timescale}
\end{table*}

The ring's maximum propagation radius increases when the target galaxy's mass grows. For the target galaxies with masses of $1.072\times 10^{10}\msun$ (model $moderate1$) and $1.072\times 10^{11}\msun$ (model $f6$), the rings can propagate out to $3h$ and $4h$, respectively. Hence a ring vanishes at a smaller radius when the disc is more dominated by MOND gravity. The existing times for rings in the above models from low-mass to massive are $14\Myr~(0.2\tdyn)$, $50\Myr~(0.9\tdyn)$ and $113\Myr~(1.5\tdyn)$. Thus the existing time scale is much shorter when the target galaxy is a low-mass dwarf disc galaxy, which indicates a deep MOND regime. The smaller vanishing radius and shorter existing time scale imply that CRGs are harder to observe in low-mass disc target galaxies. 

\begin{figure*}
  \includegraphics[width=190mm]{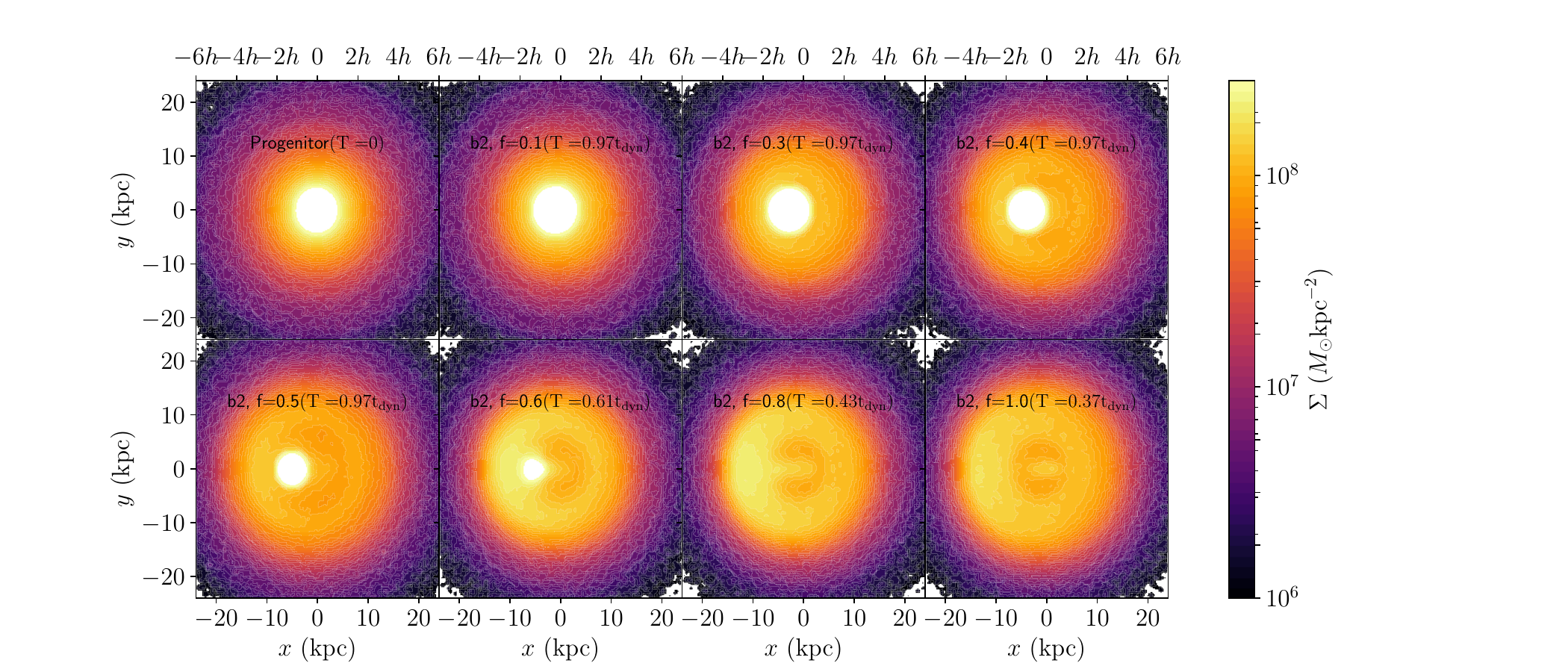}
  \caption{The projected density maps of the disc planes undergone off-centre collisions ($b2$-$set$ in Table \ref{orbpara}) with different values of $f$. 
  }  \label{projden_b2}
\end{figure*}

In another set of simulations with $f=1.0$, the results remain consistent with the previous conclusions. The maximum propagation radii of the rings are $3h$, $5.5h$ and $6.5h$ for models $deep2$, $moderate2$ and $f10$, and the existing time scales are $57\Myr~(0.9\tdyn)$, $85\Myr~(1.6\tdyn)$ and $137\Myr~(1.8\tdyn)$, corresponding to from low-mass to massive target galaxies. We summarise the time scales for the first rings of some collision models in Table \ref{timescale}. A ring on a lower-mass target galaxy disappears at a smaller radius, and the existing time scale of the ring is shorter. Moreover, given that the collisions cannot be minor ($f>0.6$) for the CRGs formed in the low-mass disc target galaxies where deep-MOND gravity is dominant, the existence of CRGs is more challenging. This is significantly different from Newtonian gravity. In Newtonian gravity, the critical ITMR is only $0.1$, and a minor head-on or off-centre collision can easily generate a ring. Since Newtonian gravity is linear, the target galaxy's overall mass and scale radius, including the dark matter halo, can be normalised to unity. As a result, the formation and propagation of rings are irrelevant to the target galaxy's mass and scale radius. In contrast, in MOND, the modified Poisson's equation is non-linear, and thus cannot be normalised for models with different masses and scale radii. A lower-mass and smaller-scale radius model emerges with stronger additional gravity than a massive, larger-scale radius model. The additional gravity, which is the so-called ``MOND effect'', stabilises the disc plane and restrains the formation and propagation of a ring structure. Hence a ring structure is harder to generate in a low-mass disc galaxy through a collision. Only collisions with an intruder galaxy mass of $\ge 50\%$ mass of a massive target galaxy, or an intruder galaxy mass of $\ge 60\%$ mass of a low-mass target galaxy can form a ring. The more difficulties in generating a ring in deeper MOND imply a lower spatial number density of CRGs in the low-mass regimes of target galaxies. 

The second rings are fainter structures compared to the first rings, as shown in the azimuthally averaged surface density profiles of the second rings in the moderate-MOND (magenta curves in the upper panels of Fig. \ref{secondary}) and the deep-MOND (black curves) regimes. For all the MOND models, the second rings are thinner structures. In general, the contrasts between the peaks and the valleys ahead of the peaks in the surface density profiles are lower compared to the first rings. 

Moreover, the values of velocity dispersion profiles for all the MOND models are about a factor of $2-3$ larger than those of azimuthal velocity profiles. Since the successful CRG models are major collision models, the target galaxies are significantly dynamically heated after the collision. 

\subsection{The stellar mass ratios of ring galaxy pairs}
Most of the observed stellar mass ratios of ring galaxy pairs are in the range of 0.1 and 1.0 \citep[e.g.,][]{Higdon_Wallin1997,Wong+2006,Romano+2008,Parker+2015,Wong+2017}. Only a few ring galaxies have an ITMR lower than 0.1 or higher than 1.0. For instance, NGC 922 contains a ring galaxy with a stellar mass of $5.22\times 10^9\msun$. The stellar mass of its companion galaxy is $2.82\times 10^8\msun$ \citep{Wong+2006}. Hence the stellar ITMR for this system is less than $0.1$. To explain the formation of such a ring galaxy with a collisional scenario, a larger overall ITMR including the dark matter halos for both galaxies is assumed  \citep[e.g.,][]{Wong+2006}. In \citet{Wong+2006}, the overall $f=0.2$, with more dark matter assembled in the companion galaxy, is used to reproduce the observed ring structure. In Newtonian dynamics, it is possible to add more mass into the dark matter halo for the companion galaxy. However, this is impossible in MOND. The dynamical mass of a galaxy in MOND is $100\%$ determined by the distribution of baryonic matter including stars and gas. A sharp question has now arisen as to whether can MOND reproduce these observed ring galaxies with a low stellar ITMR. 

As aforenoted, a collision with an ITMR of less than 0.4 cannot form a ring structure and the ring cannot propagate to a large radius such as $3h$ when $f\in[0.4, ~0.6]$ in MOND. It is difficult for dry minor collisions to cause the ring structures of the NGC 922-like galaxies. Other mechanisms should be introduced. First of all, such very rare ring galaxies might have an internal, secular origin \citep{Romero-Gomez+2007}. For instance, the ring galaxy NGC 922  is in the process of assembly and the observed ``ring'' could actually be a Fourier m=1 mode component \citep{Block+2001}. Instead of a tidal origin, such a strongly lopsided component in disc galaxies is more likely to be induced by cosmological gas accretion on the disc plane \citep{Bournaud+2005}. The secular evolution-induced ring-like structure can exist for a longer time scale, compared to the collision-induced scenario.

Another possible promising mechanism is that the rings form through other interactions, such as galaxy mergers. The observations by \citet{Elagali+2018} and \citet{Martinez-Delgado+2023} found complex tidal HI and stellar streams around NGC 922, which is not connected with the previously suggested companion dwarf galaxy \citep{Wong+2006} that induces the ring. Thus the above new observation points to a merger origin for the NGC 922-like galaxies.   
The galactic scale structures formed in MOND differ from those in Newtonian dynamics. For instance, the time scale of a dissipationless merger of galaxies is longer in MOND \citep{Nipoti+2007b}, and thus the merger frequency is lower \citep{Tiret_Combes2008}. The stellar streams, which are stripped from the satellites and might form a ring-like structure in galaxies similar to NGC 922, could potentially have a longer lifetime in the context of MOND. However, this hypothesis warrants further investigation. Systematic models for a merger origin of the ring-like structures will be studied in our follow-up projects.

Moreover, one might speculate that the intruder galaxy loses a large fraction of its mass during the collision. In a collision that forms a ring structure in the target galaxy, the low-mass companion galaxy may be disrupted or a large fraction of mass is stripped out by the strong tidal shocks \citep{Struck1999,Foster+2014,Martinez-Delgado+2023}. In MOND, the Jacobi radius is $\approx D\left(\frac{M_{\rm i}}{2M_{\rm d}}\right)^{1/3}$ \citep{ZT2006}, which is larger than that in Newtonian dynamics. Here $D$ is the distance between the satellite and the central galaxy, $M_{\rm i}$ and $M_{\rm d}$ are the baryonic masses of the companion and the target galaxies, respectively. The external field effect takes the place of the tidal field effect within the framework of MOND due to the violation of the SEP \citep{Milgrom1983a}. The companion galaxy might have already lost a large fraction of mass during the collision. We have computed the mass loss of the companion galaxy for models in Table. \ref{orbpara}, and found that the fraction is less than $5\%$ for all models when rings are at $3h$. The mass bound to the intruder galaxy is defined as stellar particles within $3r_{\rm P}$. In this radius, over $90\%$ mass is enclosed. Considering the low surface brightness of the companion galaxies, the mass cut at $3r_{\rm P}$ is reasonable.
 In addition, the suggested companion seems unlikely to suffer from strong ram-pressure stripping such that a huge amount of gas is expelled, since the NGC 922 is a low-mass galaxy \citep{Martinez-Delgado+2023}. Thus the external field shocks and ram-pressure stripping from the target galaxy are not strong enough to strip out a large amount of mass from the companion. However, in our above simulations, compact models for the intruder galaxy are used, which leads to a low fraction of mass loss by the external field stripping. If we expand the size of the intruder galaxy with a factor of $\sqrt{\frac{f}{0.1}}$, the fraction of mass loss is $5\%-7\%$ for the moderate MOND models. More sophisticated models for the intruder galaxies should be taken into account for this speculation, and we shall not focus on this issue in this work.


\section{Off-centre collisions}\label{offcentre}
\subsection{Different ITMRs}

The head-on collisions are simplified models to investigate the formation conditions for CRGs in Milgromian dynamics. However, the head-on assumption is too idealised to be a reality. It is unknown whether an off-centre collision may lead to the formation of a ring galaxy, and thus examining the formation of CRGs on off-centre orbits is important. We set up two sets of off-centre collisions on orbits with different values of impact parameter $b$, the $b1-set$ and $b2-set$ in Table \ref{orbpara}, corresponding to $b=0.1h=0.4\kpc$ and $b=0.2h=0.8\kpc$ for the mild-MOND galaxy models. In each set of off-centre collisions, the ITMR evolves from 0.1 to 1.0,  the same as those in the head-on collisions. 

We investigate the final products from the two sets of off-centre collisions and display the projected density maps of the target galaxies for the $b2-set$ if a ring forms and propagates out to a diameter of $6h$ on a disc plane in the galactocentric coordinates in Fig. \ref{projden_b2}. Here a diameter of $6h$ for the position of a ring is considered instead of a radius of $3h$ since there is an offset between the geometric centres of ring structures and the densest mass centre of galaxies. No rings form for off-centre collision models with an ITMR lower than $0.4$. We show the projected density maps of the failed CRGs at the time scale of $\approx 95 \Myr~(1.2\tdyn)$, the same time scale when a distinct ring structure evolves to a diameter of $6h$ for the model with $f=0.6$. There tends to be an obscure off-centre ring structure propagating to the diameter of $6h$ in a collision model with $f=0.5$. However, the ring structure is much more distinct when the ITMR increases to 0.6. The densest mass centre of a galaxy is more offset to the geometric centre of the ring when $f$ increases. For a model with $f\ge 0.8$, the dense central structure of the target galaxy is disrupted, and the stellar particles attracted by the intruder galaxy thicken one side of the ring. Thus the CRG appears to be an almost empty ring galaxy. The very faint central structures in the $f\ge 0.8$ models are not central nuclei but are stellar streams dragged by the gravitation of the massive intruder galaxies and projected on the disc planes. 
The rings are thick and massive for models with $f\ge 0.8$, consistent with the head-on collision results. In addition, the propagation time scale to the diameter of $6h$ is shorter as the ITMR increases. In conclusion, a ring may form through an off-centre collision in the mild-MOND models when $f\ge 0.5$. 

\begin{figure}
  \includegraphics[width=100mm]{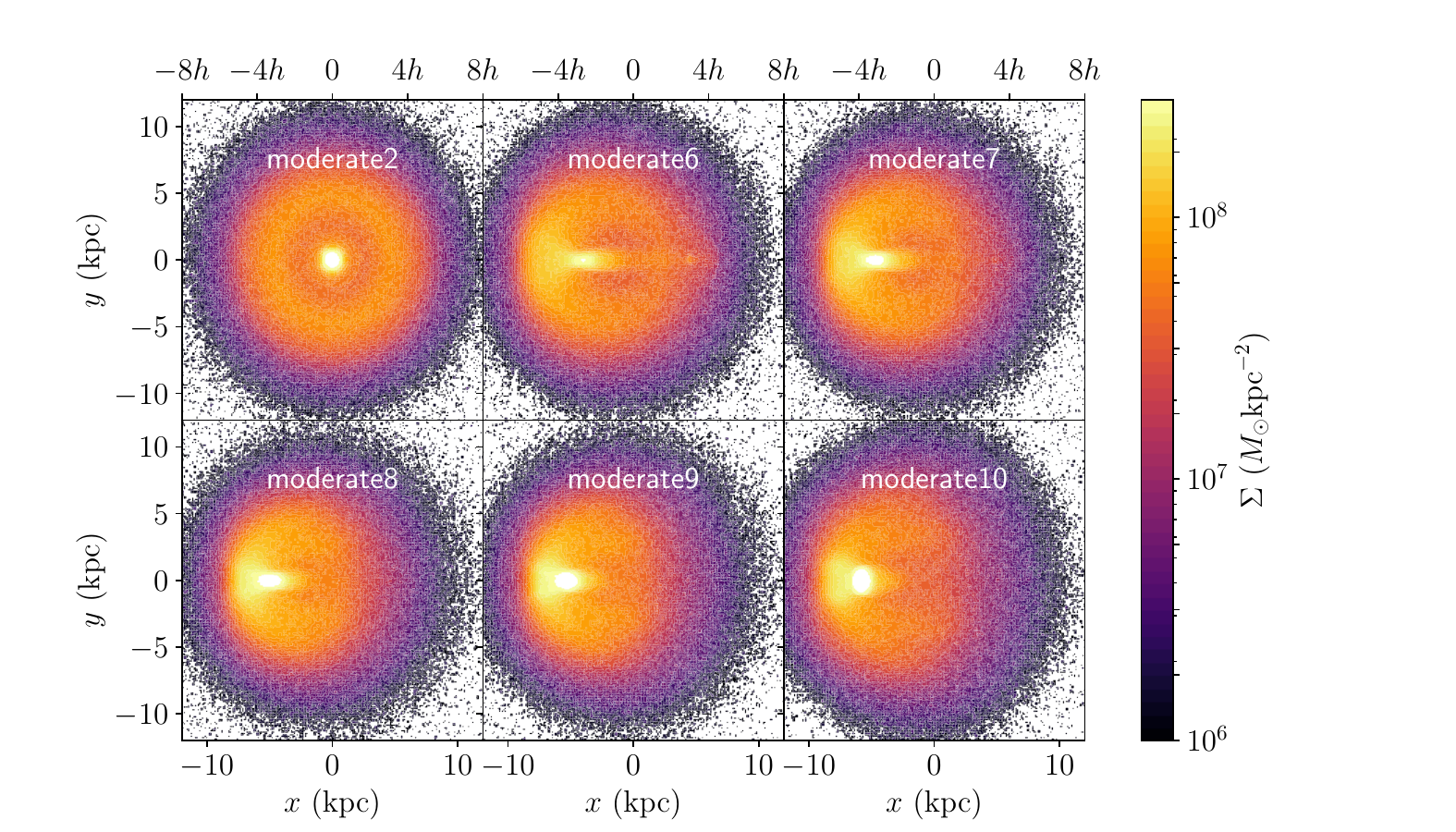}
  \caption{The projected density maps of the target disc planes at a time scale of $0.8\tdyn$ after collisions. These collisions involve a target galaxy with a mass of $1.072\times 10^{10}\msun$ and a fixed value of $f=1.0$. The impact parameter $b$ is varied in these collisions to identify the maximal value that allows the formation of a ring. 
  }  \label{b_moderate}
\end{figure}

\subsection{Impact parameter}
Now it is clear that an off-centre collision can form a ring. For a massive target model, the critical value of ITMR in off-centre collisions is the same as that in head-on collisions. A question arises as to how large the impact parameter, $b$, can form a ring. A set of simulations is performed here to examine the maximal value of $b$ in CRGs in MOND, the $moderate6$-$moderate10$ models in Table \ref{orbpara}. The moderate-MOND model for the target galaxy is adopted here. 

We show the ring structures formed on orbits with different values of $b$ in Fig. \ref{b_moderate}. The masses and values of ITMR of the three example galaxies are the same. The central nucleus and the geometry centre of a ring-like structure are more offset as the value of $b$ grows. Interestingly, different to the case of the mild-MOND model, in the $moderate2$ model, where $b=0.2h$, the central nucleus is not buried within the position of the offset ring. The stronger additional gravitation generated by the more diffused stellar density in the moderate-MOND regime may explain this. Due to the stronger additional gravitation, the disc plane of the moderate-MOND target galaxy is more stable compared to that of the massive target galaxy. Hence the perturbation from an equal-mass collision with the same scaled impact parameter ($b=0.2h$) makes it harder to destroy the disc plane of the moderate-MOND target galaxy.

Moreover, when $b\ge 0.9h$, no completed ring structures appear on the disc planes. The structures formed through these off-centre collisions are C-shaped structures.  The masses of the rings or the C-shaped structures are listed in Table \ref{orbpara}. Since the gravitational perturbation from the intruder galaxy to the target through the collision becomes weaker when the impact parameter increases, the mass of a ring or a C-shaped structure propagates out to the diameter of $6h$ decreases. The same trend is for the masses of rings or C-shaped structures propagating out to the diameter of $10h$. The ring mass of the $moderate2$ model is slightly smaller than that of models $moderate6$ and $moderate7$, because some outgoing stellar particles of the dense central nucleus overlap the position of rings in the latter models.

\section{Conclusions and discussions}\label{conclusions}
Due to the violation of SEP in MOND, the dynamical mass of the intruder galaxy keeps reducing while it is approaching the target galaxy in an orbit of head-on or off-centre collision. Thus the perturbation induced by the collision is weakened compared to that in a Newtonian intruder galaxy with a dark matter halo. It is expected harder to form a CRG in MOND. In this work, we performed a series of head-on and off-centre simulations by tuning the values of ITMR in galaxy collisions to examine the formation of CRGs.

We find that minor collisions do not form ring structures in MOND. In the mild-MOND regime, the critical ITMR is 0.5 to form a CRG and propagate out to $3h$. In general, the ring structure can propagate to a larger radius as the value of ITMR increases. Moreover, a second ring emerges after the first ring collapses back to the disc centre. The second ring is less clear compared to the first ring, and the existing time scale is longer. The time scale to observe the ring structure including the second rings is about $500 \Myr\approx 6.5\tdyn$. The mild-MOND models can reproduce CRGs that originate from massive progenitors, such as Arp 147 \citep {Romano+2008,Fogarty+2011}.

For the low-mass CRGs, their progenitors are in moderate- or deep-MOND gravity. Since the distribution of phantom dark matter follows that of the baryonic matter, the disc is more stable in MOND. In our simulations, 
we find that CRGs are harder to generate in the deeper MOND regimes. The critical ITMR is about 0.6 for the moderate- and deep-MOND models to form a ring. Since ring galaxies like the NGC 922 are observed to associate with a low-mass companion galaxy, it is difficult for MOND to explain the formation of the ring by a minor collision. To form such an observed ring structure in the NGC 922-like galaxies, one needs to introduce other mechanisms such as secular evolution \citep{Block+2001} or mergers \citep{Martinez-Delgado+2023}. After a collision, the intruder galaxy is moving outwards the target galaxy. The external field induced by the target galaxy becomes weaker, and as a result, the gravitational potential of the intruder galaxy is deepened in the outgoing orbit. Stars are hard to escape from such systems. Thus a large fraction of mass loss seems unlikely to be the mechanism for the observed NGC 922 and its low-mass companion. We have simply adopted a set of diffuse low-mass companion galaxy models in collisions. While it is possible to reproduce a ring with $f\ge 0.6$, the diffuse intruder galaxy models lose about $5\%-7\%$ masses. More sophisticated models for the intruder galaxy need to be further studied in MOND to understand the origin of the NGC 922-like galaxies.

The low-stellar-mass ring galaxies, such as the NGC 922 and Kathryn's Wheel, are dominated by moderate-MOND or deep-MOND gravity. A large amount of HI gas has been observed in these ring galaxies. Due to viscosity, gas forms a ring structure that falls behind the stellar ring and star formation is induced during the interaction with the companion galaxy \citep{Gerber+1996,Higdon1995}. In this work, we do not focus on gas dynamics, since gravity plays a much more significant role in the formation of CRGs. The conditions to form a CRG are more strict in MOND. 
The larger values of the critical ITMR to form CRGs in MOND imply a smaller space density of the CRGs than in Newtonian dynamics. 

Head-on collisions serve as simplified models to investigate the formation conditions for CRGs within the framework of MOND. Unlike head-on collisions, off-centre collisions involve non-idealized scenarios that are closer to reality. We have performed a series of off-centre collisions on orbits with different impact parameters. For the mild-MOND models, CRGs with an off-centre nucleus form with an ITMR of $\ge 0.5$. The critical ITMR is the same as that in the head-on collisions, while the morphology of the ring structure displays differences: the ring is more fuzzy and the central nucleus is more off-set as ITMR increases. When $f\ge 0.8$, the central nucleus is so offset that it overlaps the position of the ring. Thus the CRG appears to be an empty ring galaxy. The maximal impact parameter that can produce a close ring structure is $b\le 0.9h$ in our simulations. For models with an impact parameter larger than $0.9h$, a C-shaped structure forms after a major collision. The fact that a large impact parameter leads to an incompleted ring in MOND is similar to that in Newtonian simulations \citep[e.g.,][]{Mapelli_Mayer2012}.



\section*{Acknowledgements}
XW is financially supported by the Natural Science Foundation of China (Number NSFC-12073026, NSFC-11421303) and ``the Fundamental Research Funds for the Central Universities''. 

XW motivated and designed the project. LM and XW performed the simulations and analysed the results. XW and LM wrote the manuscript.

\section*{Data Availability}
The data underlying this article will be shared on reasonable request to the corresponding author.

\bibliographystyle{mnras}
\bibliography{ref.bib} 


\appendix
\section{Stability test: an example}\label{stability}
\begin{figure}
  \includegraphics[width=90mm]{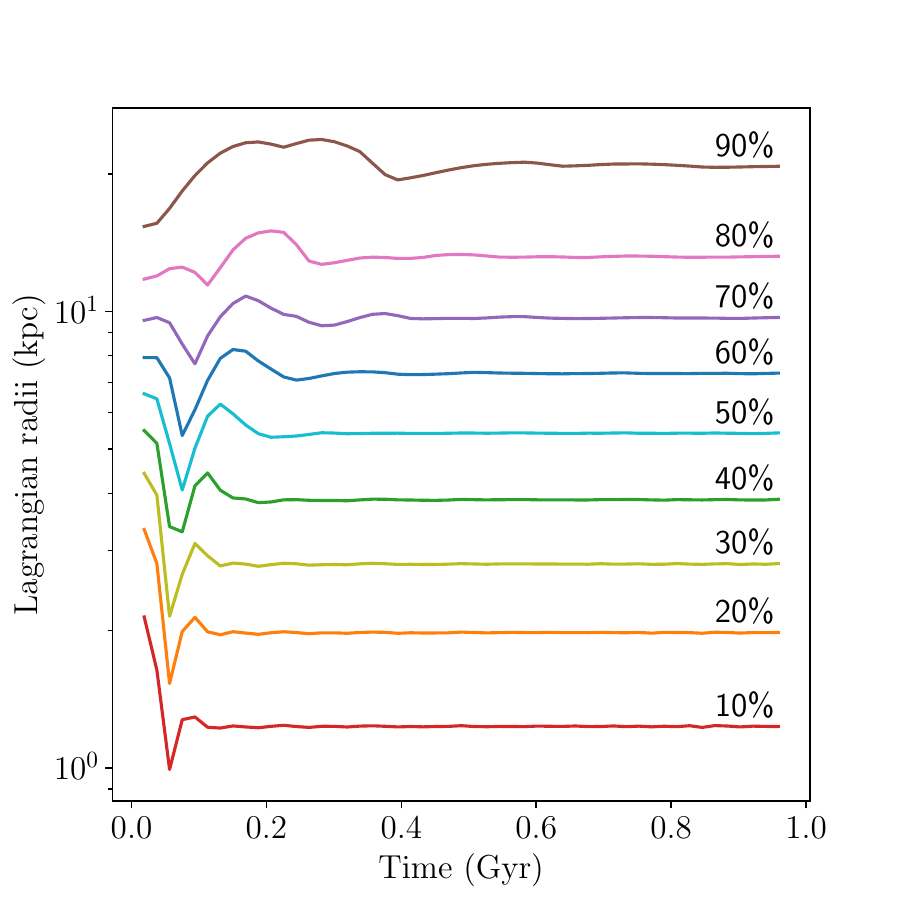}
  \caption{The figure presents the temporal evolution of the Lagrangian radii for model T1 in a stability test.
  }\label{LR}
\end{figure}

The disc galaxy models freely evolve for $1~\Gyr$ using the {\it PoR} code. The Lagrangian radii of a model are defined as the radii where $10\%,~20\%,...,~90\%$ mass are enclosed, denoted as $r_{10\%},~r_{20\%},...,~r_{90\%}$. The free evolution time scale, $1~\Gyr$, is long enough for a model to relax to a stable state in QUMOND gravity. We show the temporal evolution of Lagrangian radii of the $T1$ model (Table \ref{models}) in Fig. \ref{LR}. The Lagrangian radii of the $T1$ model exhibit a relaxation within a few dynamical times at first. The inner $r_{50\%}$ regions collapse to smaller radii, and the Lagrangian radii curves are fairly flat afterwards. For instance, after a few dynamical times, $r_{10\%}$ reduces from $\approx 2.0~\kpc$ to $1.1~\kpc$ and does not display significant evolution later. For the outer regions, the Lagrangian radii oscillate around the initial radii at the beginning. When the system is stable, the new Lagrangian radii in the outer regions such as $r_{60\%}, ~r_{70\%}$ are almost the same as the initial values.

\bsp	
\label{lastpage}
\end{document}